\documentclass[12pt]{article}
\begin{document}

\title{{\bf Statistical Evidence Against 
 \\Simple Forms of Wavefunction Collapse}
\thanks{Alberta-Thy-01-11, arXiv:yymm.nnnn [hep-th]}}

\author{
Don N. Page
\thanks{Internet address:
profdonpage@gmail.com}
\\
Theoretical Physics Institute\\
Department of Physics, University of Alberta\\
Room 238 CEB, 11322 -- 89 Avenue\\
Edmonton, Alberta, Canada T6G 2G7
}

\date{2011 January 4}

\maketitle
\large
\begin{abstract}
\baselineskip 25 pt

If the initial quantum state of the universe is a multiverse
superposition over many different sets of values of the effective
coupling `constants' of physics, and if this quantum state collapses to
an eigenstate of the set of coupling `constants' with a probability
purely proportional to the absolute square of the amplitude (with no
additional factor for something like life or consciousness), then one
should not expect that the coupling `constants' would be so biophilic as
they are observed to be.  Therefore, the observed biophilic values
(apparent fine tuning) of the coupling `constants' is statistical
evidence against such simple forms of wavefunction collapse.

\end{abstract}

\normalsize

\baselineskip 17.6 pt

\newpage

\section{Introduction}

Two alternative types of quantum theory are Copenhagen-type versions in
which the wavefunction is said to collapse at measurements, and
Everett-type versions in which the wavefunction never collapses. Often
it is believed that these two types are merely matters of interpretation
that have no observational consequences.  I have previously demonstrated
that this is not necessarily so \cite{MWI1,MWI2,Economist}.  Now I
further show that the biophilic fine-tuning observed for the `constants'
of nature in our part of the universe is statistical evidence against
certain simple collapse versions of quantum theory in an initial quantum
state that is a multiverse superposition over many different sets of the
effective coupling `constants' of physics.

Consider multiverse theories (such as what one might expect from a
string/M theory landscape \cite{BP,KKLT,Sussbook}) in which the quantum
state gives a superposition of different possible `constants' of
physics.  In Everett versions all components of the quantum state would
have real existence, and the normalized measure (`probability') for an
observation would be given by the quantum amplitudes and some
as-yet-undetermined rule for getting the measure from the quantum state
(perhaps as a normalized expectation value of a positive operator,
though I have shown it cannot be a projection operator as one might have
expected from the simplest interpretation of Born's rule
\cite{PageJCAP08,Born1,Born2,Born3,Born4}).  For a suitable rule, one
would expect that the measure would be weighted toward components of the
quantum state in which there are more observers, or a higher density of
observers, favoring components in which the `constants' have biophilic
values, which is indeed what we observe.

However, for simple collapse versions of quantum theory, one would
expect the probability of any result from the collapse of the
wavefunction to depend purely on the absolute square of the amplitude. 
There is no reason to expect this to favor biophilic values of the
`constants.' Of course, if the collapse occurred to a component of the
wavefunction with absolutely no observations, this result would not be
observed.  Therefore, to get normalized probabilities for observations,
we must condition on the occurrence of at least one observation. 
However, since a component of the wavefunction in which the universe is
sufficiently large may have some observations even if the `constants'
are not particularly biophilic, one would not expect even this selection
to restrict to components of the wavefunction in which the `constants'
have highly biophilic values, just biophilic enough to permit at least
one observation. One would then expect that most of the conditional
probabilities for observations (conditional on there being an
observation) in collapse versions of quantum theory to give values of
the `constants' that are much less biophilic than what we observe.

Therefore, the highly biophilic values that we observe for the
`constants' of physics give statistical evidence against the collapse of
the wavefunction.

\section{Consequences for toy models of multiverse states}

Let us examine the difference between collapse and no-collapse versions
of quantum theory with some simple toy models.  By a collapse (or
Copenhagen) version, I essentially mean what might more accurately be
called a single-history version, in which quantum theory gives
probabilities for various alternative sequences of events, but only one
sequence actually occurs.  Each such alternative sequence might be
called a ``history'' or a ``world.''  In contrast, by a no-collapse (or
Everett) version, I mean a `many-worlds' version, in which all of the
possible histories or worlds with nonzero quantum probabilities actually
occur, with the quantum probabilities being not propensities for
potential histories to be actualized (since all with nonzero quantum
amplitude are actual), but instead essentially measures for the
magnitudes of the existence of the various histories.

For Model I, for simplicity let there be only two alternative worlds or
sequences of events, with not-so-biophilic World 1 having quantum
measure $p_1 = 1-10^{-30}$ and $N_1 = 10^{10}$ observations, and with
highly biophilic World 2 having quantum measure $p_2 = 10^{-30}$ and
$N_2 = 10^{90}$ observations.  Assume that the number of observations in
each world is determined by the values of effective coupling `constants'
of physics that are observed in the respective observations, so that the
observations themselves show whether the world is not so biophilic or
highly biophilic.  The question then arises as to which type of
observation is more probable.

In collapse versions of quantum theory in which the probability of the
collapse of the wavefunction to each world is the quantum measure,
almost certainly (probability $1-10^{-30}$) the reduction of the quantum
state would give the not-so-biophilic World 1 and its observed effective
coupling `constants.'  On the other hand, in no-collapse versions in
which the total measure for all observations is proportional to the
number of observations as well as to the quantum measure for the
corresponding world, most of the total measure for observations would
occur for the highly biophilic World 2 and its observed effective
coupling `constants.'  That is, these simple collapse versions would
predict most probably the not-so-biophilic parameter values, whereas the
no-collapse version would predict most probably the more biophilic
parameter values.

The collapse or single-history version of quantum theory is like
assigning lottery tickets to World 1 and World 2 in the ratio $p_1:
p_2$.  Then it is as if a lottery ticket were chosen at random to select
which world, and its observations, exist.

The no-collapse or many-worlds version of quantum theory is like
assigning lottery tickets to each observation in World 1 and 2 with
ratio $p_1: p_2$, so that the ratio of the total number of lottery
tickets in World 1 to that in World 2 is $N_1 p_1: N_2 p_2$.  All the
observations exist, but with different measures for their reality,
analogous to holding different numbers of lottery tickets. Choosing a
measure-weighted observation at random is analogous to choosing a
lottery ticket at random.  The choice is not actually made (since all
observations really exist in the many-worlds version), but for assigning
probabilities to the observations despite the determinism of the
no-collapse version of quantum theory, it is helpful to imagine such a
random choice.

Let us now go to a more realistic, but still highly idealized, Model II
in which we consider the multiverse variation of one single effective
coupling `constant,' the cosmological constant $\Lambda$.  Using here
and henceforth Planck units ($\hbar = c = G = 1$), the observed value of
the cosmological constant in our World or part of the multiverse is
extremely tiny, $\Lambda_O \sim 3.5\times 10^{-122}$, but string
landscape considerations \cite{BP,KKLT,Sussbook} suggest that it could
have a huge number of different values in different parts of a
multiverse.  Presumably it could have a magnitude at least comparable to
the Planck value for either sign.

For simplicity, let us idealize the `discretuum' of values suggested by
the string landscape to a continuum and hypothesize a quantum measure
that gives a normal distribution for the cosmological constant with
standard deviation one Planck unit, so $p(\Lambda)d\Lambda =
\exp{(-0.5\Lambda^2)}d\Lambda/\sqrt{2\pi}$.  The details of this are not
important, but only the fact that the quantum measure is normalizable
and is nearly flat for values of $\Lambda$ comparable to the tiny
observed value $\Lambda_O$.

Now Martel, Shapiro, and Weinberg \cite{MSW}, following upon previous
ideas of Weinberg \cite{Wein1,Wein2,Wein3}, have shown that the
``collapsed fraction'' of matter (here by gravitational collapse, not by
the collapse of the wavefunction) to form potential observers is a very
sensitive function of $\Lambda$ that decreases rapidly if $\Lambda$ is
much larger than $\Lambda_O$.  Let us therefore consider an idealized
model in which the number of observations is proportional to
$\exp{[-(\Lambda/\Lambda_O)^2]}$.  This is of course only a very crude
hypothesis for what the actual dependence on the cosmological constant
might be, but since it is nearly flat for very small values of $\Lambda$
and falls off rapidly for $|\Lambda| \gg |\Lambda_O|$, as Martel,
Shapiro, and Weinberg found for the collapsed fraction, it will be
sufficient for our purposes.

For an observation to occur at all, we need to restrict to worlds in
which there is at least one observation, so we need a numerical
coefficient for the hypothesized gaussian dependence upon the
cosmological constant.  If the universe had infinite size (which might
be the most reasonable hypothesis), this coefficient would be infinite,
so that no matter how large $\Lambda$ were and how small the gaussian
factor were, there would still be observations somewhere in the infinite
universe.  Therefore, in simple collapse versions of quantum theory,
there would be no restriction on the cosmological constant, and its
observed probability distribution would be given purely by the quantum
measure.  If that were the normal distribution given above, the
probability that the observed value would be as small as $\Lambda_O$
would be very nearly $2\Lambda_O/\sqrt{2\pi} \sim 3\times 10^{-122}$. 
This probability is so tiny that it is very strong statistical evidence
against the hypothesis that there is wavefunction collapse to an
eigenstate of the cosmological constant, with the probabilities given by
the absolute squares of the quantum amplitudes, from an initial quantum
state that is a nearly uniform superposition of infinite universes with
the cosmological constant taking values over a range comparable to the
Planck value.

For a more conservative estimate of the probability of our observation
of the tiny value of the cosmological constant, let us suppose that the
universe is finite.  A plausible (though still very highly uncertain)
estimate of a finite size would be the size to which the universe would
inflate during slow-roll inflation from an inflaton that starts near the
Planck density.  If the inflaton were a massive scalar field,
observations of the fluctuations of the cosmic microwave background give
$m \sim 1.5\times 10^{-6}$ \cite{Lindebook,LL}.  Then if the inflation
starts with a symmetric bounce on a round three-sphere at density
$\rho_0 = 0.5 m^2 \phi_0^2$, the volume at the end of classical
slow-roll inflation is approximately \cite{Symbounce}
$[0.09644/(m\rho_0^2)]\exp{(12\pi\rho_0/m^2)} \sim \exp{(17\times
10^{12}\rho_0)}$, which would be $\sim \exp{(17\times 10^{12})}$ if the
initial density were the Planck density.

We have very little idea of the spatial density of observations after
inflation, but let us suppose that the absolute value of the logarithm
of that density is much less than $17\times 10^{12}$.  For example, if
the density were one observation per Hubble volume in our World today,
or $\sim 10^{-184}$ in Planck units, the logarithm would be $\sim
-423.7$, which is indeed negligible in comparison with $17\times
10^{12}$.  So let us hypothesize that the number of observations is,
very crudely, $\exp{[17\times 10^{12}-(\Lambda/\Lambda_O)^2]}$.  This is
greater than unity for $|\Lambda/\Lambda_O| < 4\times 10^6$, so this
model would suggest that the collapse of the wavefunction to any value
of the cosmological constant less than four million times the observed
value would still permit observations.  Therefore, if the wavefunction
for a universe inflating from the Planck density did collapse to an
eigenstate of the cosmological constant with a probability given purely
by a quantum measure that is nearly uniform for values of $\Lambda$ much
less than the Planck value, the probability plausibly would be less than
one in a million for observing our observed tiny value.  Even this
estimate of the probability is conservatively high because of the
assumed very rapid drop-off in the number of observations with
$|\Lambda| \gg |\Lambda_O|$, exponentially in $(\Lambda/\Lambda_O)^2$. 
Slower drop-offs would reduce the probability significantly.

On the other hand, for a no-collapse or Everett version of quantum
theory in which the measure for observations is weighted not only by the
quantum measure for each world with a particular value of $\Lambda$ but
also by the number (or by the number density) of observations, most of
the total measure for observations would occur for $\Lambda$ of the same
order of magnitude as $\Lambda_O$, thus explaining the highly biophilic
value observed for the cosmological constant that would be an enormously
improbable fluke in a collapse version of quantum theory with the
multiverse initial quantum state assumed here.

\section{Alternative assumptions}

The very strong statistical evidence deduced above against the collapse
of the wavefunction did include a number of crucial assumptions.  One of
course is that the initial quantum state really is a multiverse state,
with a wide range of values of the effective coupling `constants,' and
with most of the quantum measure for values that are not nearly so
biophilic as what we observe.  

A second assumption is that the wavefunction collapses to an eigenstate
of the effective coupling `constants.'  If one has a very large universe
in which the wavefunction collapse leads to a spatial distribution of
different coupling `constants' (though perhaps definite in each region,
so the collapse has eliminated the local quantum uncertainty), then a
random observation chosen from this spatial distribution could still
tend to favor biophilic values of the local coupling `constants.'  One
cannot really rule out a collapse version of quantum theory without
knowing what it says about how the wavefunction collapses.  However, it
would seem simplest to assume that the collapse would lead to an
eigenstate of the effective coupling `constants,' and it is that simple
version that I have shown leads to huge statistical problems for
plausible multiverse initial quantum states.

A third assumption is that the particular world or single history
resulting from the collapse of the wavefunction has a probability given
purely by the quantum measure (e.g., the absolute square of the
amplitude).  If the collapse itself were caused by observation (e.g., by
conscious observers), then this might weight the results to favor worlds
with more observations.  However, it would seem simpler to assume that
the collapse of the wavefunction, if it occurs at all, occurs
independently of observations.

\section{Conclusions}

I have shown that if the initial quantum state of the universe is a
multiverse state with most of the quantum measure spread over values of
the effective coupling `constants' that are not particularly biophilic,
and if the wavefunction collapses to an eigenstate of these `constants'
with a probability given purely by the absolute square of the amplitude,
then the probability is very small to observe the highly biophilic
values that we in fact do observe.  Thus our observations of highly
biophilic values is strong statistical evidence against this simple form
of wavefunction collapse under the multiverse hypothesis.

This research was partially done at the headquarters of the Asia Pacific
Center of Theoretical Physics in Pohang, Korea, and was supported by
them and by travel money from the Korea Research Foundation (KRF) Grant
funded by the Korean Government (MEST) (2010-0016-422) to Sang Pyo Kim
through Kunsan National University.  Financial support was also provided
by the Natural Sciences and Engineering Council (NSERC) of Canada.



\baselineskip 14pt
 
\begin{thebibliography}{99}

\bibitem{MWI1} D.~N.~Page, ``Observational Consequences of Many-World
Quantum Theory'' (University of Alberta report Alberta-Thy-04-99, April
1, 1999), quant-ph/9904004.

\bibitem{MWI2} D.~N.~Page, ``Can Quantum Cosmology Give Observational
Consequences of Many-Worlds Quantum Theory?'' in {\em General Relativity
and Relativistic Astrophysics, Eighth Canadian Conference, Montreal,
Quebec, 1999}, eds.\ C.\ P.\ Burgess and R.\ C.\ Myers (American
Institute of Physics, Melville, New York, 1999), pp. 225-232
[arXiv:gr-qc/0001001 (lowest existing gr-qc number)].

\bibitem{Economist} M.~Antia, ``Parallel Universes:  A World Apart,''
{\em The Economist}, May 22, 1999, p.\ 145.

\bibitem{BP} R.~Bousso and J.~Polchinski, ``Quantization of Four-Form
Fluxes and Dynamical Neutralization of the Cosmological Constant,''JHEP
{\bf 0006}, 006 (2000) [arXiv:hep-th/0004134]; ``The String Theory
Landscape,'' Sci.\ Am.\ {\bf 291}, 60-69 (2004).

\bibitem{KKLT} S.~Kachru, R.~Kallosh, A.~D.~Linde, and S.~P.~Trivedi,
``De Sitter Vacua in String Theory,'' Phys.\ Rev.\ D {\bf 68}, 046005
(2003) [arXiv:hep-th/0301240].

\bibitem{Sussbook} L.~Susskind, {\em The Cosmic Landspace: String Theory
and the Illusion of Intelligent Design} (Little, Brown and Company, New
York and Boston, 2006).

\bibitem{PageJCAP08} D.~N.~Page, ``Cosmological Measures without Volume
Weighting,'' JCAP {\bf 0810}, 025 (2008) [arXiv:0808.0351 [hep-th]].

\bibitem{Born1} D.~N.~Page, ``Insufficiency of the Quantum State for
Deducing Observational Probabilities,'' Phys.\ Lett.\ {\bf B678}, 41-44
(2009) [arXiv:0808.0722 [hep-th]].

\bibitem{Born2} D.~N.~Page, ``The Born Rule Fails in Cosmology,'' JCAP
{\bf 0907}, 008 (2009) [arXiv:0903.4888 [hep-th]].

\bibitem{Born3} D.~N.~Page, ``Born Again,'' arXiv:0907.4152 [hep-th].

\bibitem{Born4} D.~N.~Page, ``Born's Rule is Insufficient in a Large
Universe,'' arXiv:1003.2419 [hep-th].

\bibitem{MSW} H.~Martel, P.~R.~Shapiro, and S.~Weinberg, ``Likely Values
of the Cosmological Constant,'' Astrophys.\ J.\ {\bf 492}, 29-40 (1998).

\bibitem{Wein1} S.~Weinberg, ``Anthropic Bound on the Cosmological
Constant,'' Phys.\ Rev.\ Lett.\ {\bf 59}, 2607-2610 (1987).

\bibitem{Wein2} S.~Weinberg, ``The Cosmological Constant Problem,''
Rev.\ Mod.\ Phys.\ {\bf 61}, 1-23 (1989).

\bibitem{Wein3} S.~Weinberg, ``Theories of the Cosmological Constant,''
arXiv:astro-ph/9610044.

\bibitem{Lindebook} A.~Linde, {\em Particle Physics and Inflationary
Cosmology} (Harwood Academic Publishers, Chur, Switzerland, 1990).

\bibitem{LL} A.~R.~Liddle and D.~H.~Lyth, {\em Cosmological Inflation
and Large-Scale Structure} (Cambridge University Press, Cambridge,
2000).

\bibitem{Symbounce} D.~N.~Page, ``Symmetric-Bounce Quantum State of the
Universe,'' JCAP {\bf 0909}, 026 (2009) [arXiv:0907.1893].

\end{thebibliography}
\end{document}